# Simulation study of dose enhancement in a cell due to nearby carbon and oxygen in particle radiotherapy

Jae Ik Shin, Ilsung Cho, Sungho Cho, Eun Ho Kim, Yongkeun Song, Won-Gyun Jung and SeungHoonYoo

*Division of Heavy Ion Clinical Research, Korea Institute of Radiological and Medical Science, 75, Nowon-ro, Nowon-gu, Seoul, Korea*

Dongho Shin and Se Byeong Lee

*Proton Therapy Center, National Cancer Center, 323, Ilsan-ro, Ilsandong-gu, Goyang-si, Gyeonggi-do, Korea*

Myonggeun Yoon

*Department of Radiological Science, Korea University, 145, Anam-ro, Seongbuk-gu, Seoul, Korea*

SébastianIncerti

*Université Bordeaux 1, CNRS/IN2P3, Center d'EtudesNucléaires de Bordeaux-Gradignan, 33175 Gradignan, France*

Moshi Geso

*Medical Radiation Discipline, School Medical Sciences, RMIT University, Bundoora, Victoria, Australia*

Anatoly B. Rosenfeld

*Center for Medical Radiation Physics, University of Wollongong, New South Wales 2522, Australia*

The aim of this study is to investigate the dose-deposition enhancement by alpha-particle irradiation in a cellular model using carbon and oxygen chemical compositions. A simulation study was performed to study dose enhancement due to carbon and oxygen for a human cell where Geant4 code used for the alpha-particle irradiation to the cellular phantom. The characteristic of dose enhancement in the nucleus and cytoplasm by the alpha-particle radiation was investigated based on concentrations of the carbon and oxygen compositions and was compared with those by gold and gadolinium. The results show that both the carbon and oxygen-induced dose enhancement was found to be more effective than those of gold and gadolinium. We found that the dose-enhancement effect was more dominant in the nucleus than in the cytoplasm if carbon or oxygen is uniformly distributed in a whole cell. In the condition that the added chemical composition was inserted only into the cytoplasm, the effect of the dose enhancement in nucleus becomes weak. We showed that high-stopping-power materials offer a more effective dose-enhancement efficacy and suggest that the carbon nanotubes and oxygenation are promising candidates for dose utilization as dose enhancement tools in particle therapy.

Email: yoo731@kirams.re.kr

Fax: +82-2-970-1727

# I. INTRODUCTION

The chief aim of radiation therapy is to deliver a high therapeutic dose of ionizing radiation to the tumor volume without exceeding the normal tissue tolerance levels. Much of the recent effort to attain this goal has been concentrated in two distinct categories;one is in conformation of the delivered dose to the tumor volume, and the other enhancing the sensitivity of the tumor to the therapeutic radiation. The ability to conform radiation dose to the tumor has been greatly improved with the use of intensitymodulated radiotherapy, proton therapy, modulated-arc radiotherapy, and host of other technical systems. On the other hand, dose enhancement by means of nanoparticles attracted a great deal of attention as well recently. Preferential tumor uptake of contrast agents has been utilized to improve the therapeutic ratio, specificallyby changing the lower-energy photon interaction cross-sections in the immediate vicinity of the tumor, thereby delivering a highly localized dose boost to the tumor [1,2]. This is normally done by loading the target volume with contrast agents and irradiating the target with kilovoltage x-rays, where photoelectric effects are dominant in thehigh-atomic-number (Z) contrast agents such as iodine (Z=53) and gadolinium (Z=64) providing a high probability for photon interaction by the photoelectric effect [2,3] resulting into copious low energy free radicals which will be detrimental to the DNA molecules and hence enhancing the chances of cell killing and boosting radiation sensitization. Several attempts also have been made to use gold (Z=79) nanoparticles in radiation therapy [4,5] as a replacement for the conventional contrast agents. A recent study reported enhancement of radiation effects by gold nanoparticles for superficial radiation therapy using bovine aortic endothelial cells [6]. Another recent study showed that gold nanoparticles increase the radiation sensitivity of cancer cells but not that of nonmalignant cells [7]. In these in vitro experiments, dose enhancement effects were measured using cell survival levels as an indicator for the dose enhancement.

Particle beams including protons, heavy ions and alpha particles have been used in cancer therapy for their clinical superiority in delivering high doses to the targeted volume. However, with respect to particle therapy, clear evidence of dose enhancement using nanoparticles has been lacking, as the physical processes involved differ from those in photon interaction. In particle therapy, ionization processes occur directly via particle beams, whereas in photon therapy, they are induced by the photoelectric effect.Recently, the enhanced relative biological effectiveness of proton radiotherapy using gold nanoparticles was investigated [8]; however, for particle therapy, it is necessary to investigate new materials for dose enhancement at the cellular level, including specific morphology and chemical compositions.

Another candidate for dose enhancement is Carbon nanotubes (CNTs). CNTs display several properties that make them promising candidates for minimally invasive photo thermal therapy for cancer [13]. CNTs are non-toxic in normal cells and can easily be combined with cells, which means that CNTs can be used for targeting therapy that employs radiation treatment. However, there is as yet no available report on the cellular-level radiation effects of CNTs.

The oxygen enhancement also is an important factor in radiation biology, because hypoxic cells are resistant to radiation therapy. Recently, a study identified the mechanism of human and mouse breast cancer stem cell's radiation therapy resistance as the lower levels of reactive oxygen species (ROS) relative to their progeny [14]. ROS are highly reactive molecules that include oxygen ions, free radicals, and peroxides. They form as a natural byproduct of the normal metabolism of oxygen and the result of radiochemical process by ionizing radiation passing through the matter. ROS are critical intermediaries in cell killing by radiation therapy.

Understanding the biological effects of ionizing radiation on the cellular scale remains one of the challenges for current radiobiological research. During the past decade, state-of-the-art experimental irradiation facilities employing, for example, radioactive sources, classical beams and microbeams, have been established for the purpose of elucidating those effects on living cells [9]. For example, the microbeam facility of CENBG in France boasts beam resolutions on the order of a few microns in air, allowing precise control of the delivered dose – including single protons and alpha particles in the mega-electron-volt range – to targeted single cells among a cellular population. However, with this modality, experimental measurements of elementary dose deposits in cellular volumes are not possible, since the cells are maintained in the culture medium during the irradiation sequence. Incident particles of a few mega electron volts, after exiting targeted cells, are entirely absorbed by the liquid medium. Consequently, estimation of dose deposits in individual cells usually is performed by analytical calculation orby means of Monte Carlo simulation codes [10,11].

In this work, we first studied the-single-chemical composition effect on the absorbed dose in a cellular phantom. Subsequently, the beam-position effect was investigated, and then the dose enhancement effects of various chemical compositions (carbon, oxygen, gold, gadolinium)in alpha-particle irradiation were simulated for the cellular model.

## II. Material and Method

### II-1. Electronic energy loss by charged particle

In the present study, we focused on electronic stopping power. Bethe equation for the mean rate of energy loss by charged particle is

$$-\left\langle \frac{dE}{dx} \right\rangle = K z^2 \frac{Z}{A} \frac{1}{\beta^2} \left[ \frac{1}{2} \ln \frac{2 m_e c^2 \beta^2 \gamma^2 T_{max}}{I^2} - \beta^2 - \frac{\delta(\beta\gamma)}{2} \right]$$

where Z is an element with atomic number and A is mass number.

In our cellular model, the Z/A values for oxygen (0.50000), carbon (0.499542), hydrogen(0.990099), nitrogen (0.499643) and phosphorus(0.484281) were used. Additionally, the Z/A values for gold (0.401097) and gadolinium (0.406995) were used in order to add dose enhancement materials.

### II-2. Computational approach to Monte Carlo simulation

This simulation was based on the microbeam example in Geant4 (version 9.5), which simulates a realistic cellular-irradiation beamline facility configured to deliver a beam of 3 MeV alpha particles focused down to a 5 ㎛ in diameter. The cell geometry is a voxelized 3D model based on a human keratinocyte cell line [10]. The cell geometry was located at center of the volume that is a box filled with water with the size of 5 cm * 5 cm * 3 mm. Each voxel composed of the cell geometry was a box filled with the nanoparticle-aided mixture with the size of 0.36 mm * 0.36 mm * 0.16 mm.

The simulation software incorporates the physics processes of the Geant4 low-energy electromagnetic package with associated data G4LOWEM6.23 The dominant processes occurring in the experimental irradiation setup in the mega-electron-volt regime and affecting incident alpha particles are multiple scattering and ionization. For this, the Geant4 << G4EmLivemorePhysics>> constructor has been used.

The default cut-off value for the production of secondary particles was set to 10 nm, and a maximum step size limiter of 10 mm was applied in the irradiation chamber setup in order to ensure reliable tracking.

II-3. Cell compartments and added chemical compositions

Figure 1 and Table 1 show each compartment of the voxelized cellular phantom and its chemical compositions and densities. The cell cytoplasm and nucleus were based on a realistic chemical composition and a density of 1g/cm3. The mass-fraction constituents of the cytoplasm were oxygen (~58%), carbon (~20%), hydrogen (~9%), nitrogen (~8.5%), and phosphorus (~4.5%). The primary mass-fraction constituents of the nuclear material were oxygen (~74.5%) and hydrogen (~11%), with low amounts of carbon (~9%), nitrogen (~3.2%), and phosphorus (~2.6%) relative to the cytoplasm [12]. Also, a number of localized over-densities (10g/cm3) having the same material content as the nucleus were distributed throughout the cytoplasm; these substructures could represent organelles.

In this study, we investigated the carbon and oxygen on the dose deposition in the nucleus and cytoplasm using various added chemical compositions and beam positions for a targeted cell model.

**III. Result and Discussion**

   III-1. Single-chemical-composition and density effects

First, we investigated the composition and density properties of the absorbed dose in the cellular phantom which was used to maintain morphologywhile varying the chemical composition and density. For each chemical composition with the different densities, we calculated the absorbed dose deposited in the whole phantom.Figure 2(a) shows the mean total absorbed dose deposited in the phantom (at 1g/cm3 density) by each incident alpha particle with the Z/A values.In this simulation, we used $10^4$ alpha particles, about 40% of which could deposit their energy in the phantom. As the Z/A values increased, the absorbed dose deposited increased, due to the increasing stopping power. In the low-density range, the slightly different Z/A values for each component resulted in a different stopping power. With a higher stopping power, the amount of ionization could be enhanced as well. Figure 2(b) shows the density effect on the absorbed dose in the cellular phantom. We controlled the density of the phantom for each single component from 1g/cm3 to 9g/cm3. The solid line with the circular, rectangular and triangular points represents the carbon, oxygen and nitrogen. In the case of the hydrogen (Z/A=0.990099), the stopping power in hydrogen was too high; thus,at the 1g/cm3 density, the deposited absorbed dose occurred at nearly the saturated value (~0.65 Gy). As the density increased, the deposited absorbed dose at 9g/cm3, decreased to nearly 0.5 Gy. In the cases of carbon, oxygen and nitrogen, the Z/A values are almost identical (0.499542, 0.500000 and 0.499643 respectively); accordingly, the variation patterns, as shown in Fig. 2(b), alsowere very similar. The deposited absorbed dose startedat about 0.25 Gy for the 1g/cm3density, and was saturated,at 0.65 Gyfor the3g/cm3 density. High-Z materials such as gold and gadolinium result in low stopping power, and so their deposited dose and increasing slope with increasing density were lower than those of the low-Z materials.Carbon, oxygen, nitrogen and phosphorus also show a slightly decreasing deposited absorbed dose aftereach saturation point (3g/cm3 for C, O, N; 5g/cm3 for P). Theseresults can be explained by the cell's morphology and increasing mass density. The absorbed dose was calculated from the energy deposition to specific energy in Grays (J/kg). Also, as the density increased, the alpha-particle beam range decreased. Thus the area of the dose in the cell decreased as well. Figure 3 shows the dose distribution of the alpha-particle beams, which enter at the bottom of the cellular phantom and transfer in the upper direction. The range of the particle beams decreased with increasing density and stopping power.

III-2. Beam-position effect

We investigated the beam-position effect on the dose deposited in the cell-model nucleus and cytoplasm. Figures 4(a)-(d) and (e)-(h) provide a 2D projection of the mean energy deposit (in eV) per voxel per incident particle in the microbeam-irradiated cell, showing the cross-sectional plan (x,y) and the longitudinal plane(x,z), respectively. All of the plots were calculated for 104 incident alpha particles. The 2D projection data are the sum of all of the voxelvalues at each sliceof the mean energy deposit in the other direction. As the shift value increased from 0 to 30 μm in increments of 10 μm, the dose distribution position moved from the center to the left.In the case of the 0 shift, the most of the energy was deposited in the nucleus; in the 30 μm shift, contrastingly, most of the energy was deposited in the cytoplasm. Table 2 shows the maximum mean energy in the cellular phantom along with the mean absorbed dose in the nucleus and cytoplasm. The maximum mean energy deposit in the 0 shift and 30μm cases were higher than in the 10μm and 20μm shift cases, owing to the structure of the organelles,the density of which, as shown in Fig. 1 and Table 1, was 10 times that of the nucleus and cytoplasm.We also calculated the distribution of the mean absorbed dose in the nucleus and cytoplasm from each incident alpha particlein order to determinethe morphological effect on the absorbed dose. Figures 4(i)-(l) and (m)-(p) show the distribution of the mean absorbed dose in the nucleus and cytoplasm, respectively. As the shift value increased, the absorbed dose in the cytoplasm increased while the absorbed dose in the nucleus decreased. The total mean absorbed dose in the whole cellular phantom decreased as the shift value increased, owing to the morphological effect. As shift value increased, the area of the dose deposited in the cell decreased, as shown in Figs. 4(e)-(h): thus, the absorbed dose by incident particle decreased as well.

III-3. Carbon and oxygen effect

Our Monte Carlo simulationsshowed that the mean absorbed dose increased with increasing concentrations of the addedchemical composition, as shown in Fig. 5. We determined that the densities of the cytoplasm and nucleus were 1g/cm3. Since the density of the nucleus of living cells is not easily measurable and can vary according to the cell line, a density of 1 g/cm3 was considered as the reference value. The concentration of the added chemical composition varied from 0.005 g/cm3 to 0.8 g/cm3. This means that the relative concentration of the added chemical composition to the cytoplasm and nucleus varied from 0.5% to 80%. Figures 5(a) and (b) show the mean absorbed dose per particle in the 0 shift and 30 $\mu m$ shift cases, respectively. In both cases, the dose enhancement of the carbon and oxygen components was more efficient than with gold and gadolinium. In the case of the 0 shift, the absorbed dose in the nucleus was dominant for whole cellular phantom. The mean absorbed dose in the nucleus for the added carbon component case varied from 0.43 Gy (at 0.2 g/cm3) to 0.57 Gy (at 0.8g/cm3); however, the mean absorbed dose in the cytoplasm was maintained from 0.04 to 0.05 Gy. Although the added component was distributed uniformly through the whole cellular phantom, there was no distinctdose enhancement effectin the cytoplasm,becausemost of the beam's trajectory was through the nucleus. Otherwise, in the case of the 30 $\mu m$ shift, the absorbed dose in the cytoplasm was more dominant than in the nucleus. The mean absorbed dose in the cytoplasm for the added carbon component case varied from 0.10 Gy (at 0.2 g/cm3) to 0.15 Gy(at 0.8 g/cm3) with an absorbed dose in the nucleus of 0.08 Gy.

In Ref. [8], the relative biological effectiveness of proton-beam radiotherapy for prostate tumor cells with internalized gold nanoparticles was studied. Theresult represented an approximately 15-20% enhancement for Au-treated cells over untreated cells. Table 3 shows the calculated dose enhancements for two chemical compositions (C, Au) with various densities and difference. The data showan approximate maximum 38–41% enhancement for the carbon component compared with the gold component in 0.8 g/cm3 concentration, and the effect was more dominant in the nucleus than in the cytoplasm.In both the 0 and 30㎛ shift cases, as the density of the added chemical composition increased, the efficiency of the dose enhancement increased; however, the relationship between the density of added nanoparticles and toxicity needs to be investigated in an actual experiment.

Figure 5 showsthat distinct dose enhancement occurs from the 10% relative concentration of the added chemical composition (0.1 g/cm3).In our simulation study, we calculated the absorbed dose according to the chemical composition not the chemical molecules and the simulation results can be affected by the cell type, chemical composition, density, and morphology. Further study on the effect of cell type, including cancer cells and cancer stem cells is required, and this in fact is our ongoing project.

Overall, our results imply that the carbon component is more efficient than gold or gadolinium for dose enhancement in particle-based radiation therapy utilizing nanoparticles. We calculated the dose enhancement effect using chemical compositions; however, in nanoparticle-based targeting research, it is important to also investigate the effect of nanoparticle size, which implies that it is necessary to calculate dose enhancement with reference to the structureof carbon nano-materials such as CNTs. This is also one of our on-going projects.

A recent dose-enhancement experiment in vitro showed that nanoparticles were clustered in the cell's cytoplasm [7]. Other studies, such as [15],have indicated that nanoparticles were absorbed and aggregated into the cells via endocytosis. In this process, a number of gold nanoparticles are enclosed in a small pocket (vesicle) when the cell membrane is stimulated by their presence. This means that the distribution of theadded nanoparticles in the cytoplasm and nucleus cannot be uniform. In order to investigate this non-uniform distribution effect, we calculated the absorbed dose in the cytoplasm and nucleus under the condition thatthe added chemical composition was inserted only into the cytoplasm, not the nucleus or organelles.

Figures 6(a) and (b) show the mean absorbed dose in the whole cell for non-uniform-distributed added materials in the 0 ㎛ and 30 ㎛ cases, respectively.We calculated the absorbed dose with the increasing concentrations of added materials only in the cytoplasm. The concentration varied from 0.2 g/cm3 to 0.8 g/cm3. The quantitative analysis results are listed in Table 4.

In the case of the 0 shift, the gold-induced dose enhancement effect showed no response (under 1 %), and that of carbon decreasedby about a factor of 10 relative to the uniformly distributed case. For the 30 ㎛ shift, the effect was almost the same as with the uniform distribution. This result implies that the nanoparticles have to be inserted into the nucleus to inducesignificant DNA damage. Also, cell death resulting from damage to the cytoplasm needs to be investigated further. In the case in which the added materials are inserted into the only cytoplasm, the dose enhancement of the nucleus region (including DNA) becomes weaker.Kuncic et al. [16] noted that non-targeted damage (damage to thecytoplasm region including the mitochondria, lysosomes, and membranes) is markedly different from the mechanistic response to DNA-targeted radiation. Non-targeted biological responses are determined by the cell's entire state, including all proteins and macromolecules in its cytoplasm; however, when electromagnetic interactions occur primarily outside the nucleus, the ensuing biological damage is poorly understood. Several newly recognizedresponses, collectively referred to as non-targeted responses, are manifested by effects such as mutagenesis, genomic instability, the bystander effect, changes in gene expression, and even adaptive responses [17-23]. The mechanismby which enhancement of dose to the cytoplasm induces cell death (decreasing viability) requires further study.

## IV. CONCLUSION

In this work, we performed a simulation study to investigate the dose enhancement in a cellular model (human normal cells) by means of various chemical compositions. To determine the absorbed dose in the nucleus and cytoplasm, the Geant4 microbeam example (version 9.5) was implemented. We verified the relation between the absorbed dose and the stopping power with a single-chemical-composition-based phantom and investigated the chemical compositions and morphological effects of the alpha-particle irradiation according to a microbeam's position shift. Carbon and oxygen werefound to be more effective for dose enhancement than gadolinium and gold in particle irradiations. We also showed that high-stopping-power materials offer a more effective dose-enhancement efficacy;on this basis, we suggestthat the carbon nanotubes (CNTs) and oxygenation are promising candidatesfor dose utilization as dose enhancement tools in particle therapy.

**ACKNOWLEDGEMENT**

This research was supported by the National NuclearR&D Program through the National Research Foundation ofKorea(NRF) funded by the Ministry of Education, Science andTechnology (2010-0026071).Additional support was received from the National Cancer Center, Korea (Research GrantNo. 1210210-1).Travel support for S. Incerti was also provided by the International Associated Laboratory "France Korea Particle Physics Laboratory" (FKPPL)

**REREFENCES**

[1] Mello RS, Callisen H, Winter J, Kagan AR, and Norman A,Med. Phys.**10**, 75-8(1983).

[2] Mesa AV. Norman A, Solberg TD, Demareo JJ, and Smathers JB, Phys. Med. Biol.44, 1955-68 (1999).

[3] Robar JL, Roccio SA, and Martin MA, Phys. Med. Biol. **47**, 2433-49 (2002).

[4] Hainfeld JF, Stalkin DN, and Smiloitz HM, Phys. Med. Biol. **49**, 309-15(2004).

[5] Hainfeld JF, Stalkin DN, Focella TM, and Smiloitz HM, Br. J. Radiol. **79**, 248-53(2006).

[6] Kong T, Zeng J, Wang X, Yang X, Yang J, McQuarrie S, et. al., Small,**4**, 1537-43 (2008).

[7] Wan NordianaRahman, NourBishara, Trevor Ackerly, Cheng Fa He, Price Jackson, Christopher Wong, Robert Davison, and Moshi Geso, Nanomedicine,**5**, 136-42(2009).

[8] Jerimy C. Polf, Lawrence F. Bronk, Wouter H. P. Driessen, and WadihArap, Appl. Phys. Lett.**98**, 193702(2011).

[9]Incerti S, Gault N, Habchi C, Lefaix J. L, MorettoPh, Poncy J. L, PouthierTh, and Seznec H, Radiat. Prot. Dosim. **122**, 327-9(2006).

[10]S. Incerti, H. Seznec, M. Simon, Ph. Barberet, C. Habchi, and Ph. Moretto, Radiat. Prot. Dosim.**133**, 2-11(2009).

[11] Ph.Barberet, F.Vianna, M.Karamitros, T.Brun, N.Gordillo, Ph.Moretto, S.Incerti and H.Seznec,Phys. Med. Biol. **57**, 2189-207 (2012).

[12] J.-P. Alard, V. Bodez, A. Tchirkovet. al., Radiation Research, **158**, 650-6 (2002).

[13] Andrew R. Burke, Ravi N. Singh, David L. Carroll, James C.S Wood, Ralph B. D'Agostino Jr., Pulickel M. Ajayan, Frank M. Torti, and Suzy V. Torti, Biomaterials,**33**, 2961-70(2012).

[14]Maximilian Diehn, Robert W. Cho, Neethan A. Lobo, TomerKalisky, Mary Jo Dorie, Angela N. Kulp,DalongQian, Jessica S. Lam, Laurie E. Ailles, Manzhi Wong, Benzion Joshua, Michael J. Kaplan, Irene Wapnir, Frederick M. Dirbas, George Somlo, Carlos Garberoglio, Benjamin Paz, Jeannie Shen, Sean K. Lau,Stephen R. Quake, J. Martin Brown, Irving L. Weissman,and Michael F. Clarke,Nature,**458**, 780-5(2009).

[15] Tsai SW, Chen YY, andLiaw JW, Sensors,**8**, 2306-16(2008).

[16] ZdenkaKuncic, Hilary L.Byrne, Aimee L. McNamara, Susanna Guatelli, WestaDomanova, and SébastienIncerti, Computational and mathematical methods in medicine,**2012**, 147252 (2012).

[17] K. M. Prise, G. Schettino, M. Folkard, and K. D. Held, Lancet Oncology,**6**, 520-528 (2005).

[18] L.-J. Wu, G. Randers-Pehrson, A. Xu et al., Proceedings of the National Academy ofSciences of the United States of America,**96**,4959–4964(1999).

[20] C. Shao, M. Folkard, B. D. Michael, and K. M. Prise, Proceedings of the National Academy of

Sciences of the UnitedStates of America,**101**, 13495–13500(2004).

[21] K. M. Prise and J. M. O'Sullivan, Nature Reviews Cancer, **9**, 351–360 (2009).

[22] D. Averbeck, Mutation Research,**687**, 7–12(2010).

[23] E. G. Wright, Mutation Research,**687**, 28–33(2010).

[24] G. Schettino, S. T. Al Rashid, and K. M. Prise, Mutation Research,**704**, 68–77 (2010).

Fig. 1 Compartments of cell phantom; (a) cytoplasm, (b) nucleus, (c) organelles

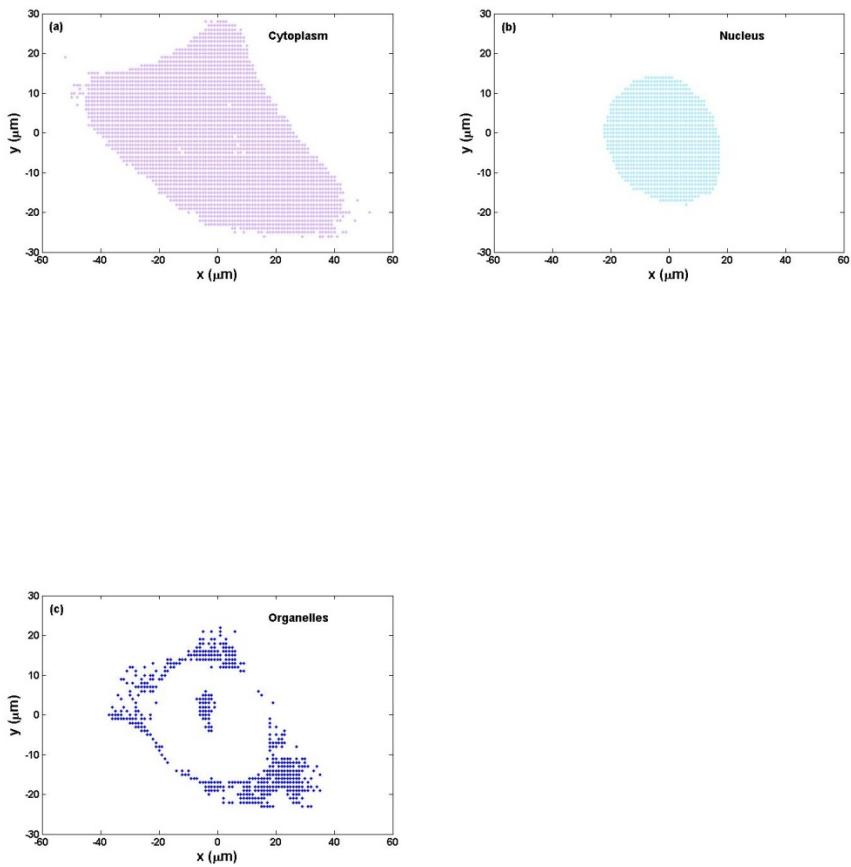

Table 1 Chemical compositions and densities of each cellular compartment

|   | Cytoplasm(1g/cm$^3$) | Nucleus(1g/cm$^3$) | Organelles(10g/cm$^3$) |
|---|---|---|---|
| H | 9% | 10.64 % | 10.64% |
| O | 58% | 74.5 % | 74.5% |
| C | 20% | 9.04 % | 9.04% |
| N | 8.5% | 3.21 % | 3.21% |
| P | 4.5% | 2.61 % | 2.61% |
| Total | 100% | 100% | 100% |

Fig. 2 Z/A effect on absorbed dose at 1g/cm$^3$ density for each chemical composition (a) and density effect on absorbed dose for each chemical composition (b) in cellular phantom

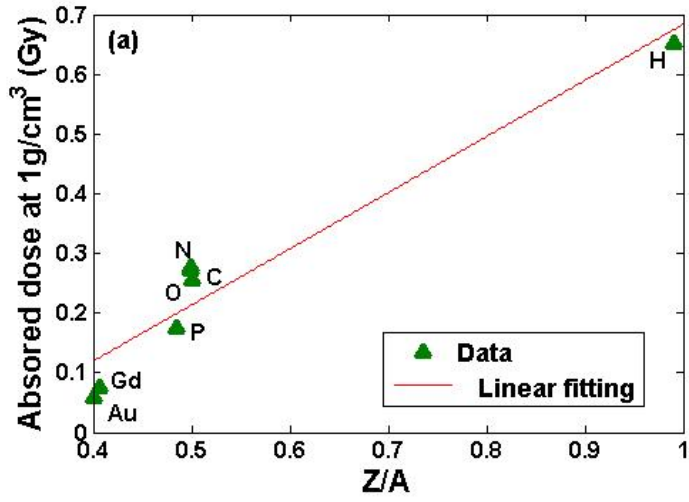

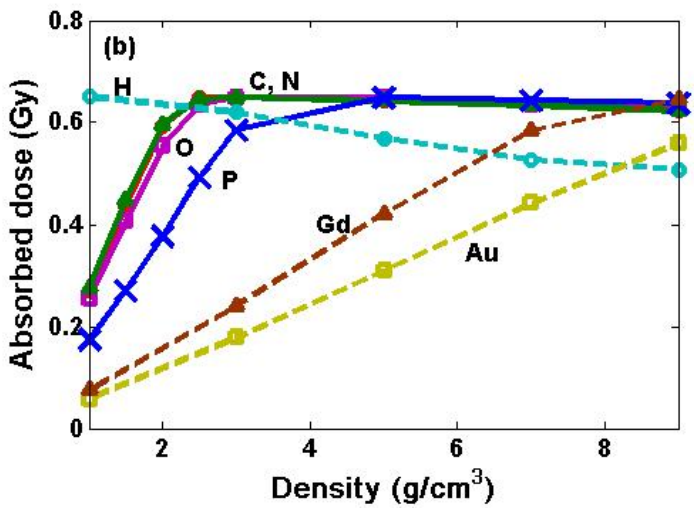

Fig. 3 Stopping power and density effects on dose distribution in single component cellular phantom (Alpha particles enter at the bottom of the phantom and transfer in the upper direction.)

(a)-(c): Dose distribution of alpha particles in Au (Z/A=0.401097) phantom ((a) 1, (b) 5 and (c) 9 g/cm$^3$)

(d)-(f): Dose distribution of alpha particles in C (Z/A=0.499542) phantom ((d) 1, (e) 5 and (f) 9 g/cm$^3$)

(g)-(i): Dose distribution of alpha particles in H (Z/A=0.990099) phantom ((g) 1, (h) 5 and (i) 9 g/cm$^3$)

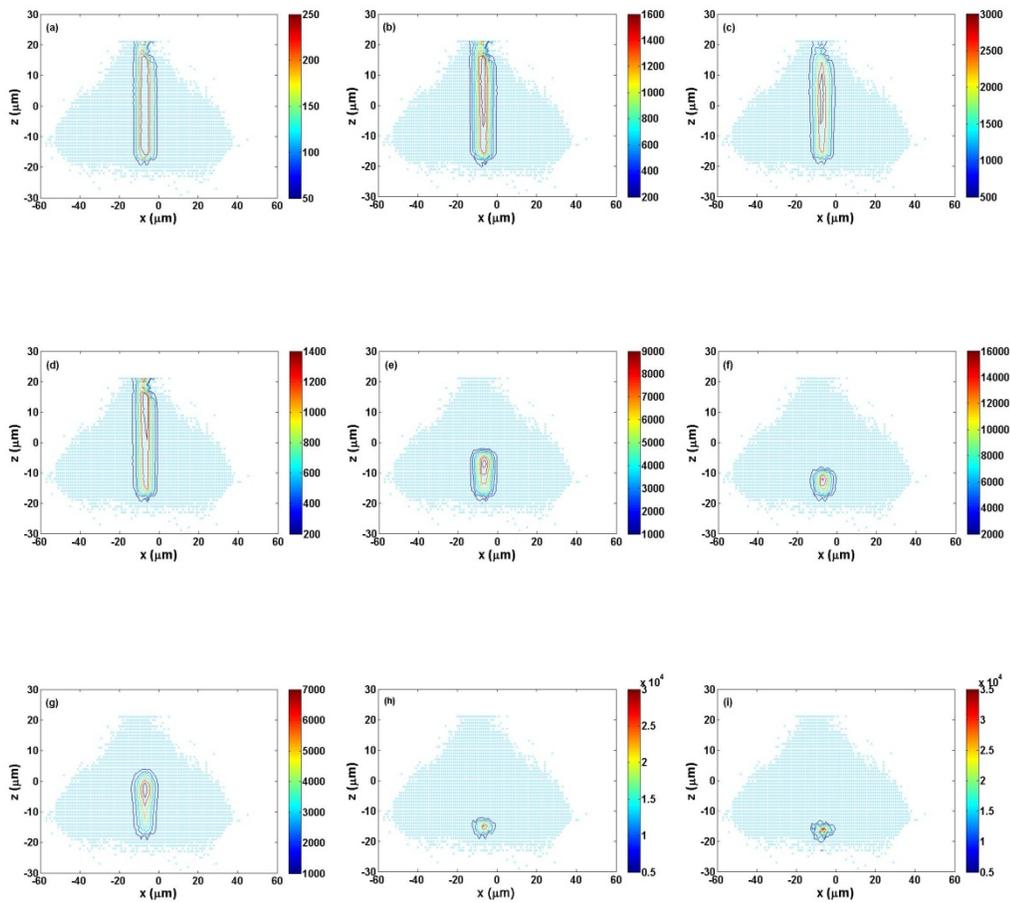

Fig. 4 Beam position effect on energy and absorbed dose deposited in cellular phantom for 0 ((a), (e), (i), (m)), 10 ((b), (f), (j), (n)), 20 ((c), (g), (k), (o)), 30 μm ((d), (h), (l), (p)) shift cases from center

(a)-(d) : 2D (x,y) projection of mean energy deposit (in eV) per voxel per incident particle in the microbeam-irradiated cell

(e)-(h) : 2D (x,z) projection of mean energy deposit (in eV) per voxel per incident particle in the microbeam-irradiated cell

(i)-(l) : Distribution of absorbed dose (in Gy) in nucleus per incident particle

(m)-(p) : Distribution of absorbed dose (in Gy) in cytoplasm per incident particle

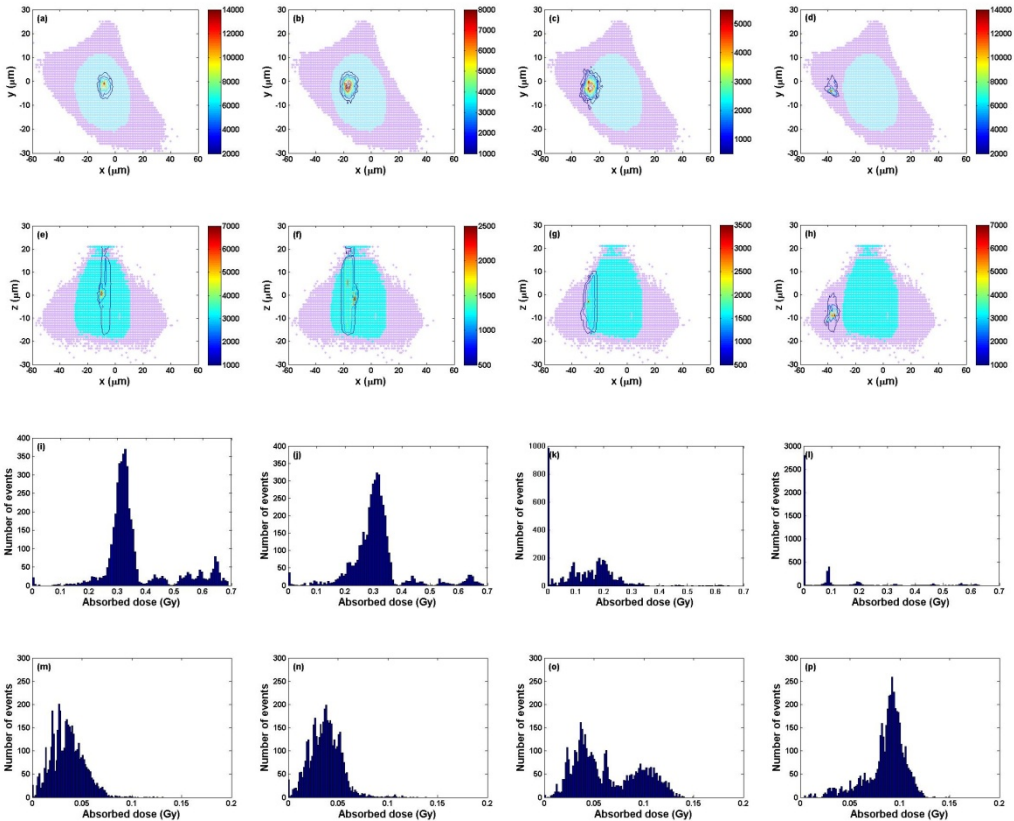

Table 2. Maximum energy in cellular phantom and mean absorbed dose in nucleus and cytoplasm for various beam positions

| | Center | 10 μm shift | 20 μm shift | 30 μm shift |
|---|---|---|---|---|

|  | (0 μm shift) |  |  |  |
| --- | --- | --- | --- | --- |
| Maximum mean energy per voxel in phantom | 3024 eV | 954.5 eV | 1983 eV | 2875 eV |
| Mean absorbed dose in nucleus | 0.36 Gy | 0.31 Gy | 0.13 Gy | 0.08 Gy |
| Mean absorbed dose in cytoplasm | 0.04 Gy | 0.04 Gy | 0.06 Gy | 0.09 Gy |
| Total mean absorbed dose in whole cell | 0.40 Gy | 0.35 Gy | 0.19 Gy | 0.17 Gy |

Fig. 5 Effect of chemical composition on dose enhancement (uniform distribution case). Mean absorbed dose in whole cell; (a) 0μm shift, (b) 30μm shift

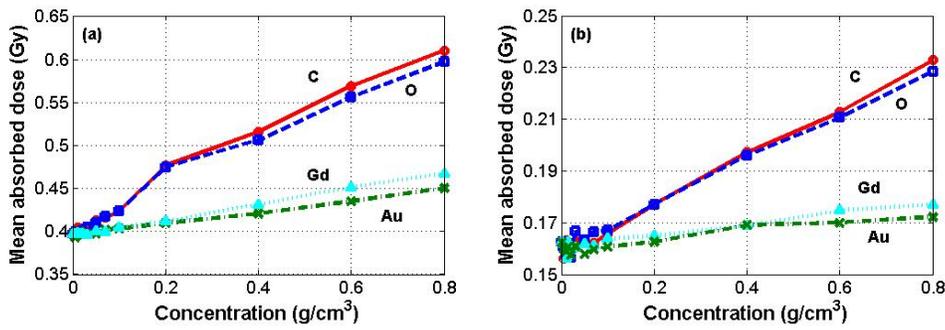

Table 3. Calculated dose enhancement for chemical compositions (C, Au) with various densities and difference (uniform distribution case)

|  | 0 shift case | | | | 30μm shift case | | | |
|---|---|---|---|---|---|---|---|---|
| Added density(g/cm3) | 0.2 | 0.4 | 0.6 | 0.8 | 0.2 | 0.4 | 0.6 | 0.8 |
| Dose enhancement for C (%) | 20.0 | 30.0 | 42.5 | 53.5 | 12.5 | 25.0 | 31.3 | 43.8 |
| Dose enhancement for Au (%) | 2.5 | 5.0 | 10.0 | 12.5 | 0.1 | 4.1 | 4.6 | 6.0 |
| Difference (%) | 17.5 | 25.0 | 32.5 | 41.0 | 12.4 | 20.9 | 26.7 | 37.8 |

Fig. 6 Effect of chemical composition on dose enhancement (non-uniform distribution case). Mean absorbed dose in whole cell; (a) 0μm shift, (b) 30μm shift

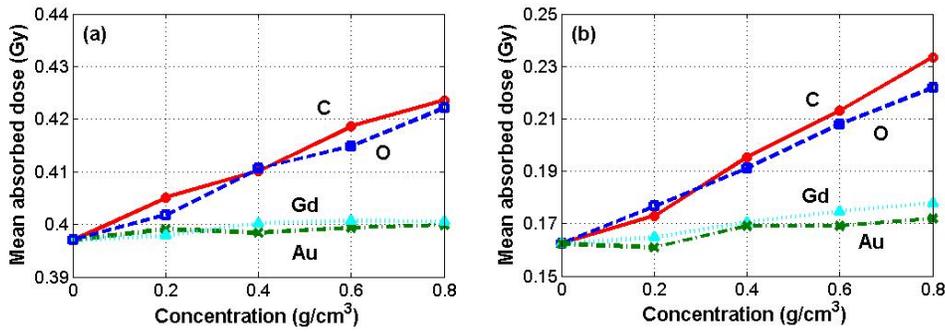

Table 4. Calculated dose enhancement for chemical compositions (C, Au) with various densities and difference (non-uniform distribution case)

|  | 0 shift case | | | | 30μm shift case | | | |
|---|---|---|---|---|---|---|---|---|
| Added density (g/cm3) | 0.2 | 0.4 | 0.6 | 0.8 | 0.2 | 0.4 | 0.6 | 0.8 |
| Dose enhancement for C (%) | 2 | 3.3 | 5.4 | 6.7 | 6.4 | 20.3 | 31.2 | 43.7 |
| Dose enhancement for Au (%) | 0.5 | 0.4 | 0.6 | 0.7 | -0.9 | 4.2 | 4.1 | 5.9 |
| Difference (%) | 1.5 | 2.9 | 4.8 | 6.0 | 7.3 | 16.1 | 27.1 | 37.8 |